\documentclass[prd,a4paper,aps,twocolumn,
               nofootinbib,nobibnotes,superscriptaddress,preprintnumbers]
              {revtex4}

\usepackage[dvips]{graphicx}
\usepackage{amsmath,amssymb,mathrsfs}
\usepackage{bm}
\usepackage{times}
\usepackage{epsfig}
\usepackage{verbatim}
\usepackage{bm}
\usepackage[utf8]{inputenc}
\usepackage{graphics}
\usepackage{graphicx,epsfig,amssymb,amsmath,color, cancel}
\usepackage{soul}

\newcommand{\be}{\begin{equation}}
\newcommand{\ee}{\end{equation}}
\newcommand{\ba}{\begin{array}}
\newcommand{\ea}{\end{array}}
\newcommand{\bea}{\begin{eqnarray}}
\newcommand{\eea}{\end{eqnarray}}

\begin{document}

\preprint{DESY 16-068}

\title{A First-Order Electroweak Phase Transition from Varying Yukawas}

\author{Iason Baldes}
\author{Thomas Konstandin}
\affiliation{DESY, Notkestra{\ss}e 85, D-22607 Hamburg, Germany}
\author{G\'eraldine Servant}
\affiliation{DESY, Notkestra{\ss}e 85, D-22607 Hamburg, Germany}
\affiliation{II. Institute of Theoretical Physics, University of Hamburg, D-22761 Hamburg, Germany}

\date{\today}

\begin{abstract}

 We show that the dynamics responsible for the variation of the Yukawa couplings of the Standard Model fermions generically leads to a very strong first-order electroweak phase transition, assuming that the Yukawa couplings are large and of order 1 before the electroweak phase transition and reach their present value afterwards.
 There are good motivations to consider that the flavour structure could emerge during electroweak symmetry breaking, for example if the Froggatt-Nielsen field dynamics were linked to the Higgs field. In this paper, we do not need to assume any particular theory of flavour and show in a model-independent way how the nature of the electroweak phase transition is completely changed when the Standard Model Yukawas vary at the same time as the Higgs is acquiring its vacuum expectation value. 
\textcolor{black}{The thermal contribution of the fermions creates a barrier between the symmetric and broken phase minima of the effective potential, leading to a first-order phase transition.}
This offers new routes for generating the baryon asymmetry at the electroweak scale, strongly tied to flavour models. 

\end{abstract}

\maketitle

\section{Introduction}

While the Higgs sector has started to be well measured at the LHC,  the nature of the electroweak phase transition (EWPT) still remains very poorly constrained. 
{In fact, it depends only weakly on the value of the Higgs mass which solely provides information about the Higgs potential in the vicinity of its broken-electroweak-symmetry minimum and not on its global properties.}
In the Standard Model (SM), the EWPT is a rapid crossover \cite{Csikor:1998eu} but minimal extensions of the SM can make it first-order. {By further constraining the Higgs couplings}, the {next} run of the LHC {will provide} new probes of models leading to first-order EWPT, which would have {major} implications for EW baryogenesis and therefore our understanding of the origin of the matter antimatter asymmetry of the universe~\cite{Cohen:1990it}.
The EW baryogenesis framework relies on the existence of a strongly first-order EW phase transition~\cite{Trodden:1998ym,Morrissey:2012db}. The baryon asymmetry is produced in the vicinity of the EW symmetry breaking bubble walls \cite{Konstandin:2013caa} where all three Sakharov conditions~\cite{Sakharov:1967dj} are at work. Particle distributions depart from thermal equilibrium and CP-violating currents are converted into baryons by sphalerons~\cite{Manton:1983nd,Klinkhamer:1984di}.
  
  A variety of mechanisms leading to a first-order EWPT have been proposed. 
  Extensions of the Standard Model (SM) where the barrier separating the symmetric and broken phase minima is thermally generated due to loops  of new bosonic modes in the Higgs effective potential, e.g. 
  in the supersymmetric extension of the SM~\cite{Katz:2014bha},
  have been severely constrained by Higgs measurements.
  In contrast, a model where the Higgs potential is modified at tree level due to the couplings of the Higgs to an additional singlet, leading to potentially a 2-step EWPT is a very simple scenario which remains difficult to test, e.g.~\cite{Espinosa:2011ax,Curtin:2014jma}.
  Finally, 
  models where EW symmetry breaking is induced by a dilaton \cite{Konstandin:2011dr,Konstandin:2011ds,Servant:2014bla}, with couplings similar to the Higgs, although relatively suppressed, are another interesting and testable possibility.

We present a different path here. We show that if the Yukawa couplings $y_{ij}$ in the interactions between the SM fermions and the Higgs boson, $ y_{ij} \overline{f}^i_L \Phi^{(c)} f^j_R$, vary during the EWPT, from a value of order 1 at the beginning of the EWPT to their present value at the end of the EWPT when $\langle \Phi \rangle=v /\sqrt{2}$, this can lead naturally to a very strong first-order PT.

\section{Emerging flavour during EW symmetry breaking}

A variation of the Yukawa couplings during the EWPT is actually well-motivated when considering 
how the flavour structure and fermion mass hierarchy of the SM may emerge.
There are three main mechanisms to describe fermion masses $m_f=y_f v/\sqrt{2}$: spontaneously broken abelian flavour symmetries as originally proposed by Froggatt and Nielsen~\cite{Froggatt:1978nt} (FN), localisation of the profiles of the fermionic zero modes in extra dimensions~\cite{Gherghetta:2000qt} and partial  fermion compositeness in composite Higgs models~\cite{Panico:2015jxa}. The last two scenarios may be related by holography~\cite{Maldacena:1997re,Gherghetta:2010cj}. 
The scale $M$ at which the flavour structure emerges is not a-priori constrained. 
In FN constructions, the Yukawa couplings  are controlled by the breaking parameter of a flavour symmetry. A scalar field $\chi$ carrying a negative unit of the abelian charge develops a vacuum expectation value (VEV) and
\begin{equation}
y_{ij} \sim (\langle \chi \rangle /M)^{-q_i+q_j+q_H},
\end{equation}
where the $q$'s are the flavor charges of the fermions and the Higgs. For an appropriate choice of flavour charges and with  $\langle \chi \rangle /M \sim 0.22$, measured masses and mixings can be well described.
In most FN constructions, the prejudice is that the scale $M$  is very high, close to the GUT or Planck scale. However, it could be lower, and even close to the EW scale~\cite{Babu:1999me,Giudice:2008uua,Tsumura:2009yf,Calibbi:2012at,Berger:2014gga,Calibbi:2015sfa,Bauer:2015fxa,Bauer:2015kzy,Bauer:2016rxs,Huitu:2016pwk}. 

While there is a huge literature on models advocated to explain the fermion masses~\cite{Babu:2009fd}, {until recently there was} no study on the associated cosmology. On the other hand, in all flavour models, Yukawa couplings are controlled by the VEV of some scalar fields  (the so-called ``flavons") and it is natural to wonder about their cosmological dynamics. Our working assumption is that the flavon couples to the Higgs and therefore the flavon and the Higgs VEV dynamics are intertwined, motivating the possibility that the Yukawas vary during the EWPT. 
{The various implications of this framework for electroweak baryogenesis {are} presented in a series of  papers (reviewed in~\cite{Servant:2018xcs}). In particular,  the CKM matrix as the unique CP-violating source is discussed 
in~\cite{Bruggisser:2017lhc}.
Specific models of varying Yukawas have been studied: Froggatt-Nielsen~\cite{Baldes:2016gaf,Baldes:2018nel}, Randall-Sundrum~\cite{vonHarling:2016vhf} and composite Higgs~\cite{Bruggisser:2018mus,Bruggisser:2018mrt}.}

In this letter, the key point we want to make is that 
{even before specifying the dynamics responsible for the evolution of the Yukawas, we can derive important implications  for  the nature of the EWPT. 
 The fact that the Yukawas of the SM were large during the EWPT is enough to change the nature of the EWPT, even when ignoring the flavon dynamics and only considering the SM degrees of freedom (dof) in the first place. We refer to the above references for specific realisations and the associated experimental constraints.}

\section{Effect of fermionic masses on the EWPT}

The physics of the effect of varying Yukawas is related to the contribution of effective relativistic dof $g_*$ to the effective potential $V_{\rm eff} \supset - g_{*}\pi^{2}T^{4}/90$. Regions in Higgs space in which species are massive correspond to a decrease in $g_*$ and hence an increase in $V_{\rm eff}$. The effect of species coupled to the Higgs is therefore to delay and hence strengthen the phase transition. In the usually assumed case where the Yukawas have the same values during the EWPT as today, all Yukawas except the one of the top quark are small and therefore almost all fermions are light even in the broken phase during the EWPT. Therefore there is no significant change in $g_*$ during the EWPT and the effect of the light fermions is negligible. Crucially, the contribution of bosonic species to the finite-$T$ effective potential also includes a term cubic in the mass and hence bosonic dof not only delay the phase transition but also create a barrier between the two minima. However, the effect of the SM bosons is insufficient to provide a strong first-order phase transition \cite{Csikor:1998eu}. Thus, the common lore consists of adding additional bosonic degrees of freedom to strengthen the phase transition. As mentioned in the introduction, this has been severely constrained at the LHC.

On the other hand, it was shown in~\cite{Carena:2004ha}
that adding new strongly-coupled fermions with {\it constant} Yukawa couplings can also help to strengthen the EWPT. Though these do not create a thermal barrier on their own, they can lead to a decrease in $g_*$ between the symmetric and broken phases and hence delay and strengthen the phase transition. However, these models suffer from a vacuum instability near the EW scale due to the strong coupling of the new fermions and new bosons are also needed to cure this instability.

In our approach of varying Yukawas, these problems are alleviated. We are interested in models where the variation of the Yukawa couplings is due to the VEV of a flavon field, coupled to the Higgs, whose VEV therefore also varies during the EWPT. If the Yukawa couplings decrease with the Higgs background value $\phi$, the SM fermions can be massless both in the symmetric phase, at $\phi=0$, as well as at $\phi \sim v$ due to the falling couplings, but be massive somewhere in between, i.e in the region $0 < \phi < v$. This raises the potential in this area and can therefore create a barrier. The quantitative size of this effect is encoded in the effective potential which we shall study below.

We stress that this does not mean that the Yukawa couplings are controlled solely by the Higgs field, i.e. the Higgs need not itself be the flavon (such a scenario is strongly constrained by various Higgs and flavour measurements, see~\cite{Giudice:2008uua,Babu:1999me,Bauer:2015fxa,Bauer:2015kzy}). The variation of the Yukawas is related to the variation of the Higgs VEV during the EWPT (during which the flavon VEV may also change) but the  Yukawas today do not depend on the Higgs VEV $v=246$ GeV nor are the Higgs-fermion couplings sizeably affected. Model-dependent implementations {are} presented elsewhere~\cite{Baldes:2016gaf,vonHarling:2016vhf,Bruggisser:2018mus,Bruggisser:2018mrt,Baldes:2018nel,Servant:2018xcs}.

\begin{figure}[t]
\begin{center}
\end{center}
\includegraphics[width=230pt]{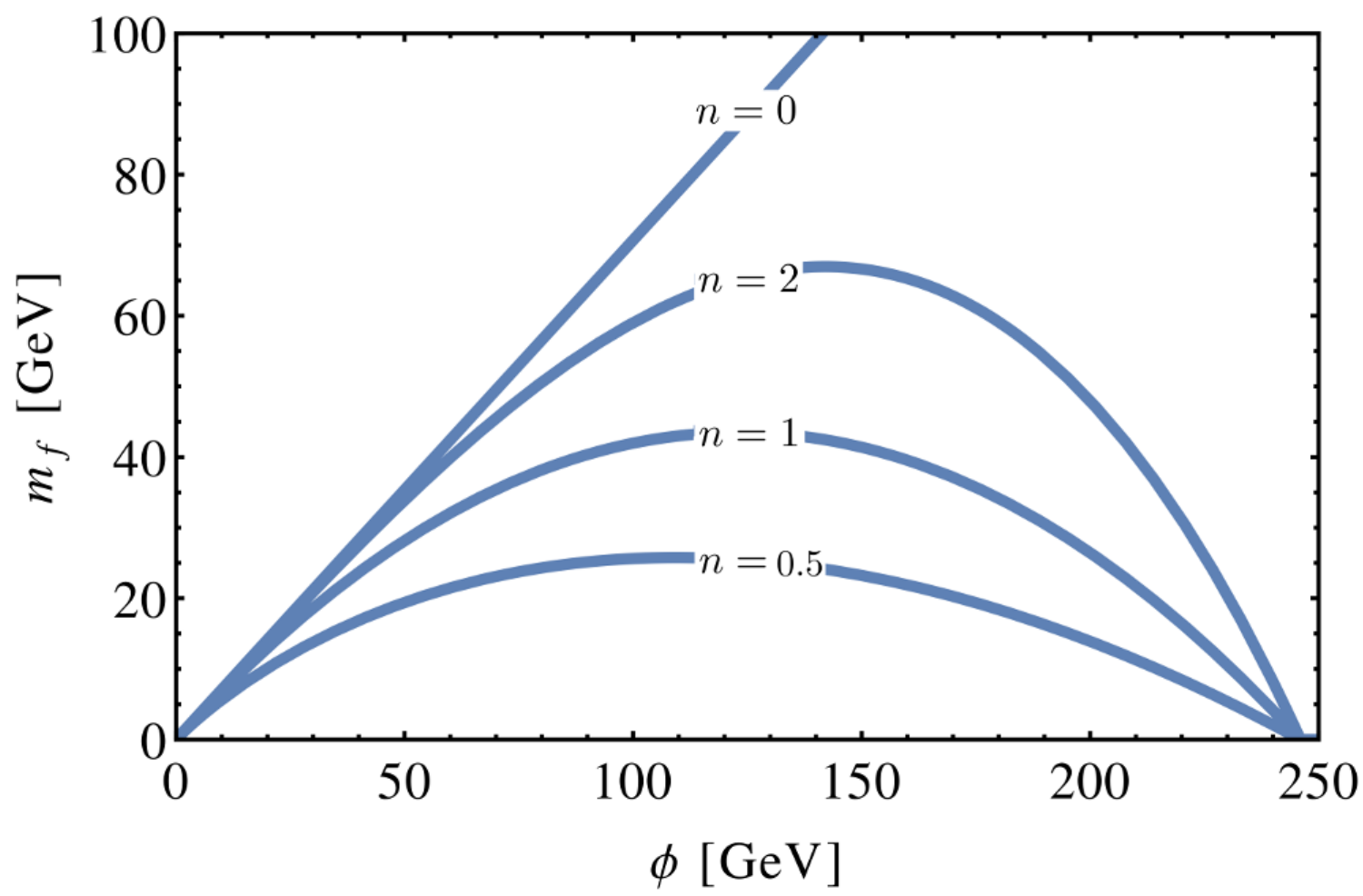}
\caption{The mass of a fermionic species as a function of $\phi$ for a constant Yukawa coupling, $n=0$, and varying Yukawas. In the constant Yukawa case we take $y(\phi)=1$. For the varying Yukawa cases we take 
$y_{1}=1$ and $y_{0}=0$ (see Eq.~\ref{eq:Yukawavsphi}).}
\label{fig:masses}
\end{figure}

The aim of this letter is  to stress the model-independent features of the physics of Yukawa variation.
We will therefore present results using the following ansatz for the variation of the Yukawa related to the variation of the Higgs VEV itself (once the path is known in the $(\phi,\chi)$ plane, we can eliminate the flavon VEV via the relation $\chi(\phi)$):
\begin{equation}
y(\phi)=
  \begin{cases}
   y_{1}\left( 1 -  \left[ \frac{\phi}{v} \right]^{n} \right)+y_{0} &  \quad \mbox{for} \quad  \phi \leq v,\\
    y_{0} & \quad  \mbox{for} \quad   \phi \geq v.\\
  \end{cases}
\label{eq:Yukawavsphi}
\end{equation}
The mass of the fermion species is given by
	\begin{equation}
	m_{f}=\frac{y(\phi)\phi}{\sqrt{2}} \ .
	\end{equation}
{and we illustrate the dependence of $m_{f}$ on $\phi$ in Fig.~\ref{fig:masses}.}
Equation~(\ref{eq:Yukawavsphi}) just expresses the fact that before the EWPT, the Yukawas take values $y_1$ and after the EWPT they take their present value $y_0$. The power $n$ is just a parametrisation of how fast the variation is taking place and is therefore encoding the model dependence. Depending on the underlying model and the dynamics of the bounce, the Higgs field variation will follow the flavon field variation at different speeds. 
{Once the path during the phase transition in field space is known, one can eliminate the flavon VEV and reduce the tunnelling problem to a one-dimensional one. Unlike in an Effective Field Theory, the parametric dependence of the Yukawa on the Higgs can be very complicated, see e.g. ~\cite{vonHarling:2016vhf}.  But since the Yukawas of the light quarks essentially drop from unity to small values, the family of functions given in ~(\ref{eq:Yukawavsphi}) is representative of what can be observed in a full numerical treatment. Our equation ~(\ref{eq:Yukawavsphi}) should be viewed as a simple toy ansatz to study the qualitative effect of Yukawa variation on the EW phase transition and not as a full realistic model.}
Large values of $n$ mean the Yukawa coupling remain large for a greater range of $\phi$ away from zero. We will see that large $n$ strengthen the phase transition.

We study the strength of the EWPT for different choices of $n$, $y_{1}\sim {\cal O}(1)$ and the number of degrees of freedom, $g$, of the species with the $\phi$-dependent Yukawa coupling. The results do not depend strongly on the choice of $y_{0}$ as long as $y_{0} \ll 1$. The top Yukawa is assumed to be constant and take its SM value. 

Of course, in a realistic model the different fermion species will take on different values of $n$, $y_{1}$ and $y_{0}$ (also the underlying model determines whether only quarks, only leptons or all fermion masses are controlled by the same flavon). Our aim here is to simply illustrate the effect through a simple ansatz and an overall variation of $n$, $g$ and $y_{1}$.

The possibility that the Yukawa couplings could change during the EWPT was raised in \cite{Berkooz:2004kx} but the impact on the nature of the EWPT was ignored, the emphasis was on the possibility to get large CP violation from the CKM matrix during the EWPT. We show in the next section the three main effects that Eq.~(\ref{eq:Yukawavsphi}) has on the Higgs effective potential.

\begin{figure}[t]
\begin{center}
\end{center}
\includegraphics[width=230pt]{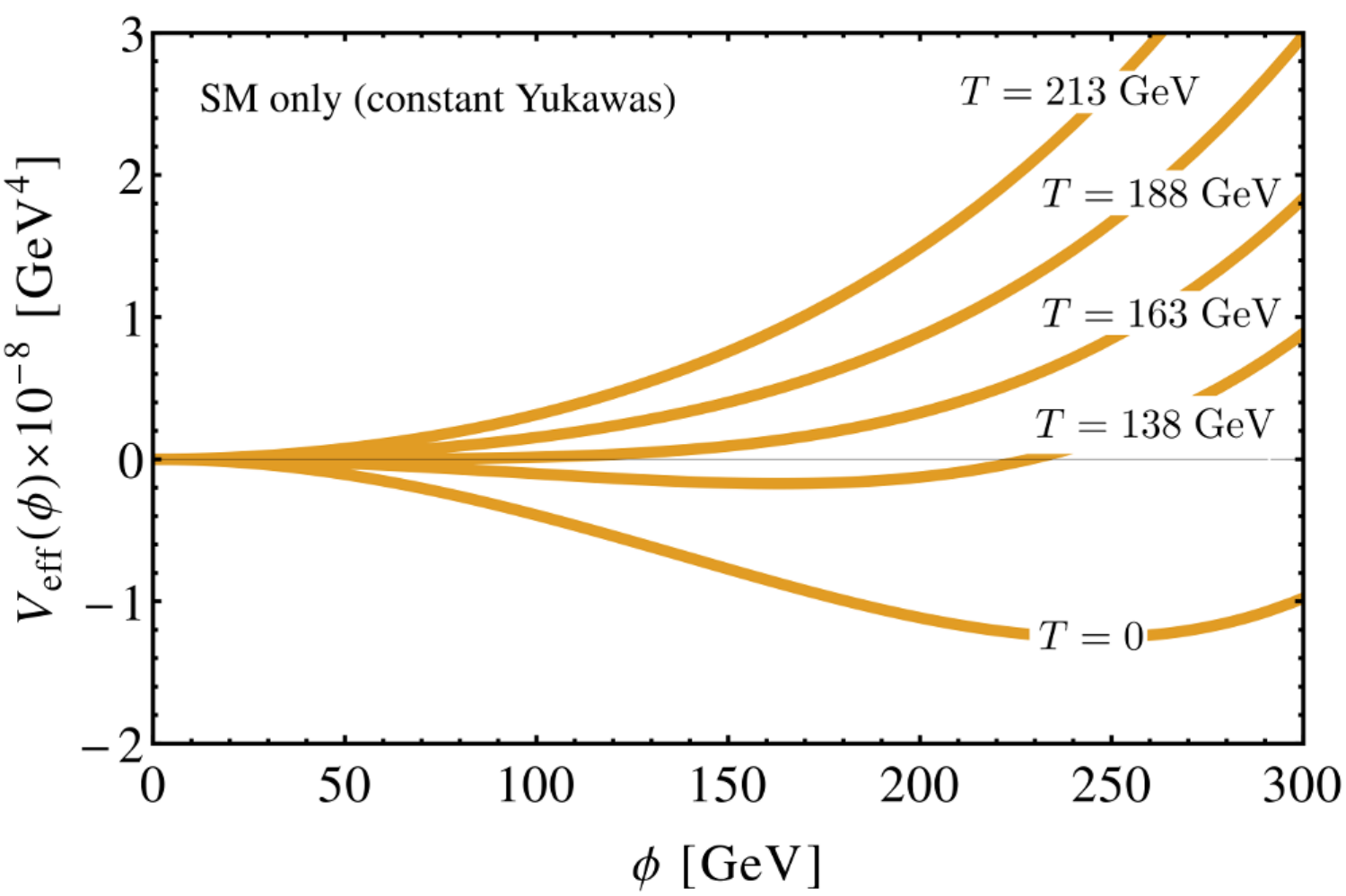}
\includegraphics[width=230pt]{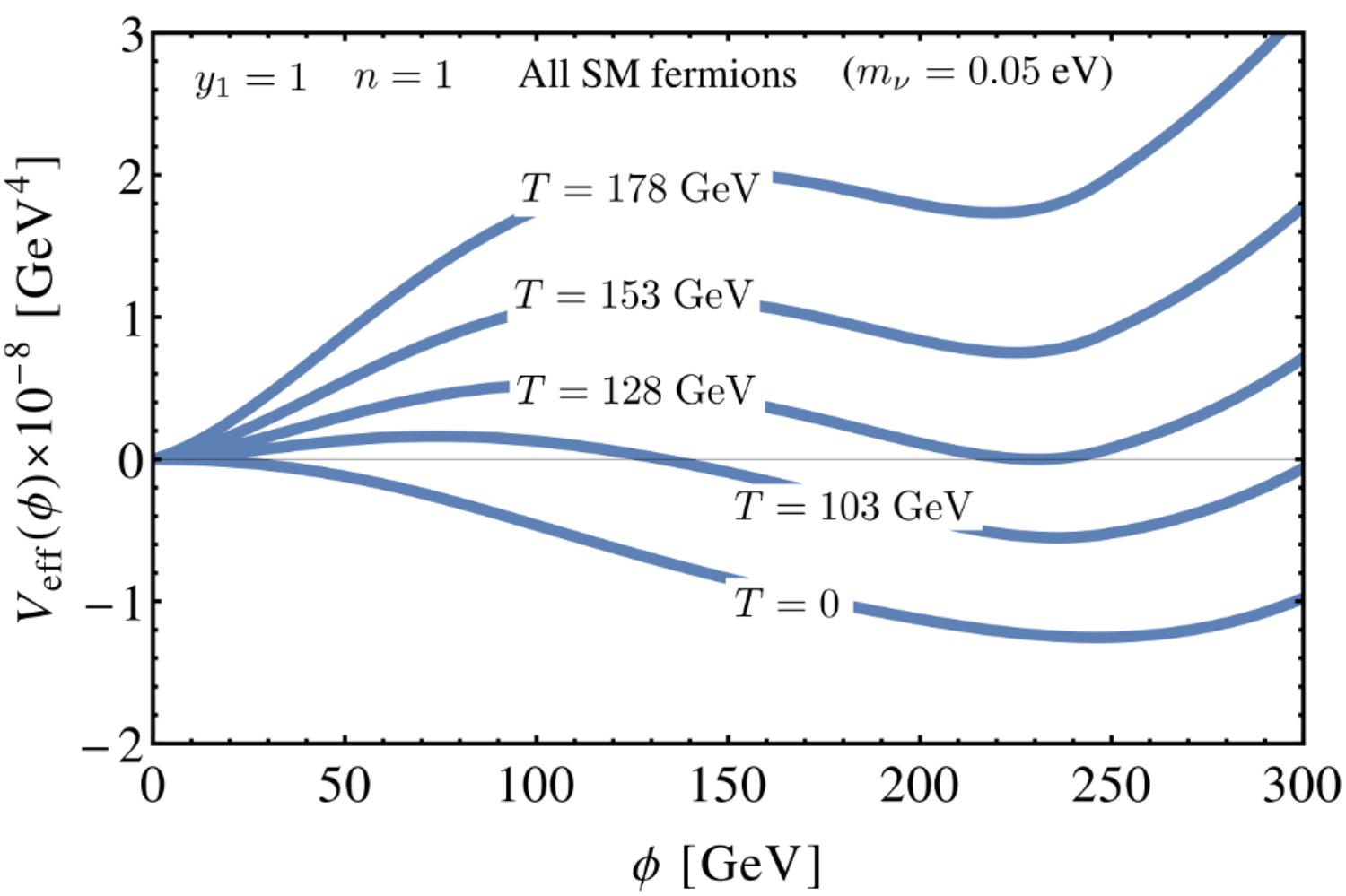}
\caption{The evolution of the effective potential with temperature in the SM (top) and with varying Yukawas 
(bottom). The varying Yukawa calculation includes all SM fermions with $y_{1}=1$, $n_{1}=1$ and their respective $y_{0}$, chosen to return the observed fermion masses today (for the neutrinos we have assumed Dirac neutrinos and $m_{\nu}=0.05$ eV). In the varying Yukawa case we find a first-order phase transition with
 $\phi_{c}=230$ GeV and $T_{c}=128$ GeV (second order transition at $T_{c}=163$ GeV for the constant Yukawa case).}
\label{fig:SM_vs_varying}
\end{figure}

\begin{figure}[t]
\begin{center}
\end{center}
\includegraphics[width=230pt]{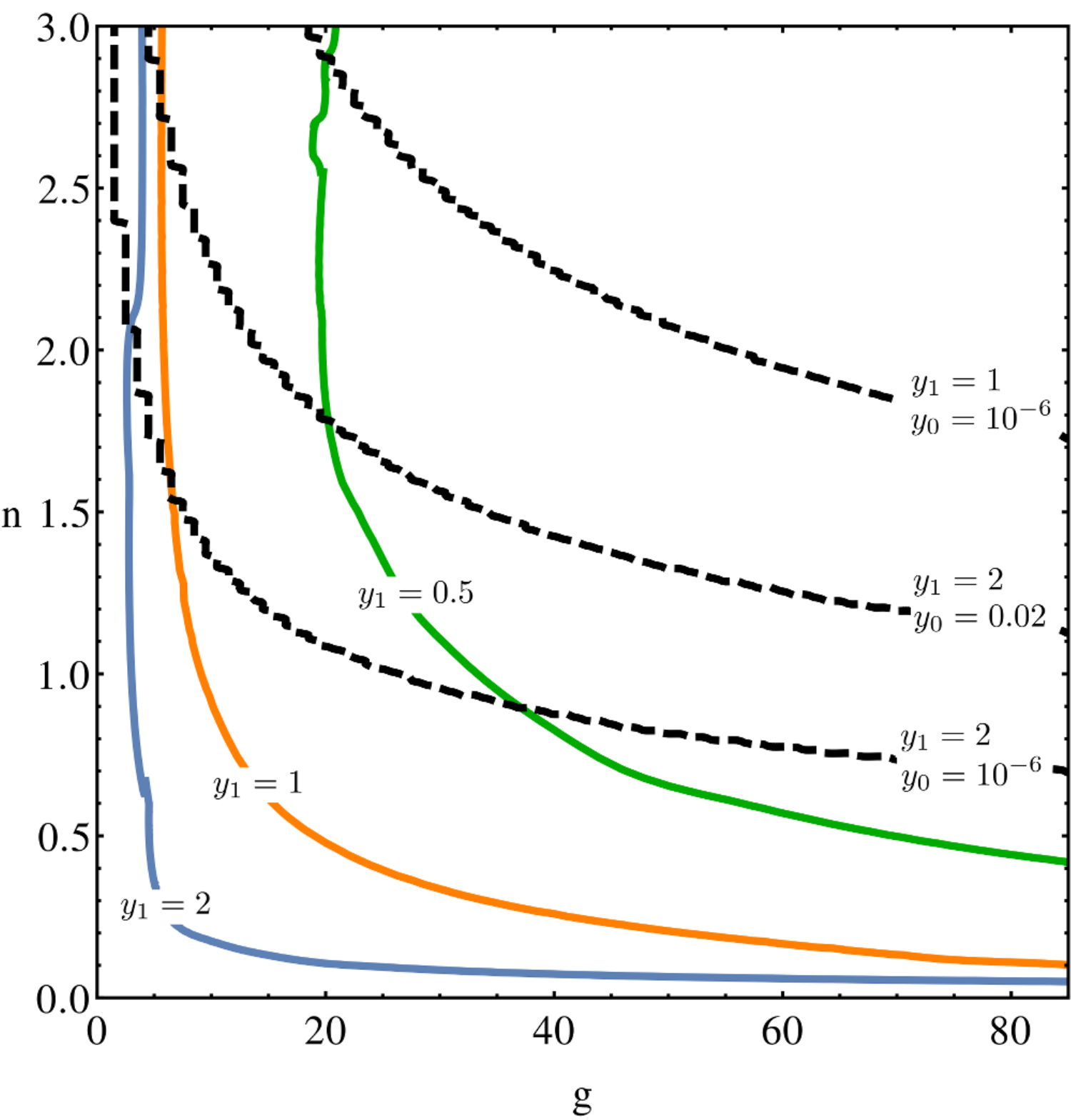}
\caption{Solid lines: Contours of $\phi_{c}/T_{c}=1$ for different choices of $y_1$ and $y_{0}=0.02$, areas above these lines allow for EW baryogenesis. Dashed lines: areas above these lines are disallowed (for the indicated choices of $y_{1}$ and $y_{0}$) due to the EW minimum not being the global one.}
\label{fig:n_againts_g}
\end{figure}

\section{Effective Higgs potential with varying Yukawas}

We consider the effective potential given by the sum of the tree level potential, the one-loop zero temperature correction, the one-loop finite temperature correction and the daisy correction~\cite{Quiros:1999jp}
	\begin{align}
	V_{\rm eff} = V_{\rm tree}(\phi)+V_{1}^{0}(\phi)+V_{1}^{T}(\phi,T)+V_{\rm Daisy}(\phi,T).	
	\end{align}
	In the framework we have in mind, this potential depends as well on the additional flavon field(s) coupling to the Higgs. However, for the generic points we want to stress, we should ignore the flavon(s) degrees of freedom and 
	take the SM tree level potential. We study the evolution of the effective potential with temperature numerically, including the SM fermionic dof with varying Yukawas, in addition to the usual bosonic SM fields. An example of the evolution of the effective potential with varying Yukawa couplings, with a comparison to the SM case (constant Yukawas), is shown in Fig.~\ref{fig:SM_vs_varying}. We next scan over $n$ and $g$ for different choices of $y_{1}$ and find the strength of the phase transition, as characterised by the ratio of the critical VEV to temperature, $\phi_{c}/T_{c}$ (successful EW baryogenesis requires $\phi_{c}/T_{c} \gtrsim 1$~\cite{Ahriche:2007jp}). Our results are summarised in Fig.~\ref{fig:n_againts_g}. Below we discuss the different terms of the effective potential and identify the contributions leading to a strong first-order phase transition.\\	
	
{\bf 1) Effects from the $T=0$ one-loop potential}:	
 The one-loop zero temperature correction is given by
	\begin{align}
	\nonumber
	V_{1}^{0}(\phi)&=&  \sum_{i} \frac{g_{i}(-1)^{F}}{64\pi^2} \bigg\{ &
								m_{i}^{4}(\phi)\left(\mathrm{Log}\left[\frac{m_{i}^{2}(\phi)}{m_{i}^{2}(v)} \right] - \frac{3}{2}\right)   \\
							&&	&   +2m_{i}^{2}(\phi)m_{i}^{2}(v)\bigg\}, 	\label{eq:oneloop}
	\end{align}
where $g_{i}$ are the SM degrees of freedom, $F=0 \; (1)$ for bosons (fermions) and we have ignored the contribution of the Goldstone bosons ($g_{i}$ does not strictly correspond to the degrees of freedom present, hence both the longitudinal gauge boson dof and the Goldstones should be summed in the Landau gauge, however, their contribution is subdominant and we therefore neglect them~\cite{Delaunay:2007wb}).  The field dependent gauge boson masses are
	$M_{W}^{2}(\phi)  = {g_{2}^{2}}\phi^{2}/4, 
	M_{Z}^{2}(\phi)  = ({g_{2}^{2}+g_{Y}^{2}})\phi^{2}/4,$
where $g_{2}$ ($g_{Y}$) is the weak isospin (hypercharge) gauge coupling. 

It is clear from (\ref{eq:oneloop}) that the effect of a large fermionic mass is to significantly lower the potential between $\phi=0$ and $\phi=v$. This can lead to weaker -- rather than stronger -- phase transitions for increasing $y_{1}$ or $n$ in some areas of parameter space. In addition, it can lead to the EW minimum no longer being the global minimum.  \textcolor{black}{Note the effect grows logarithmically as $y_{0}$ decreases.} The regions of parameter space in which the global minimum is not the EW one are shown in Fig.~\ref{fig:n_againts_g}.\\

%
%
%
\begin{figure}[t]
\begin{center}
\includegraphics[width=230pt]{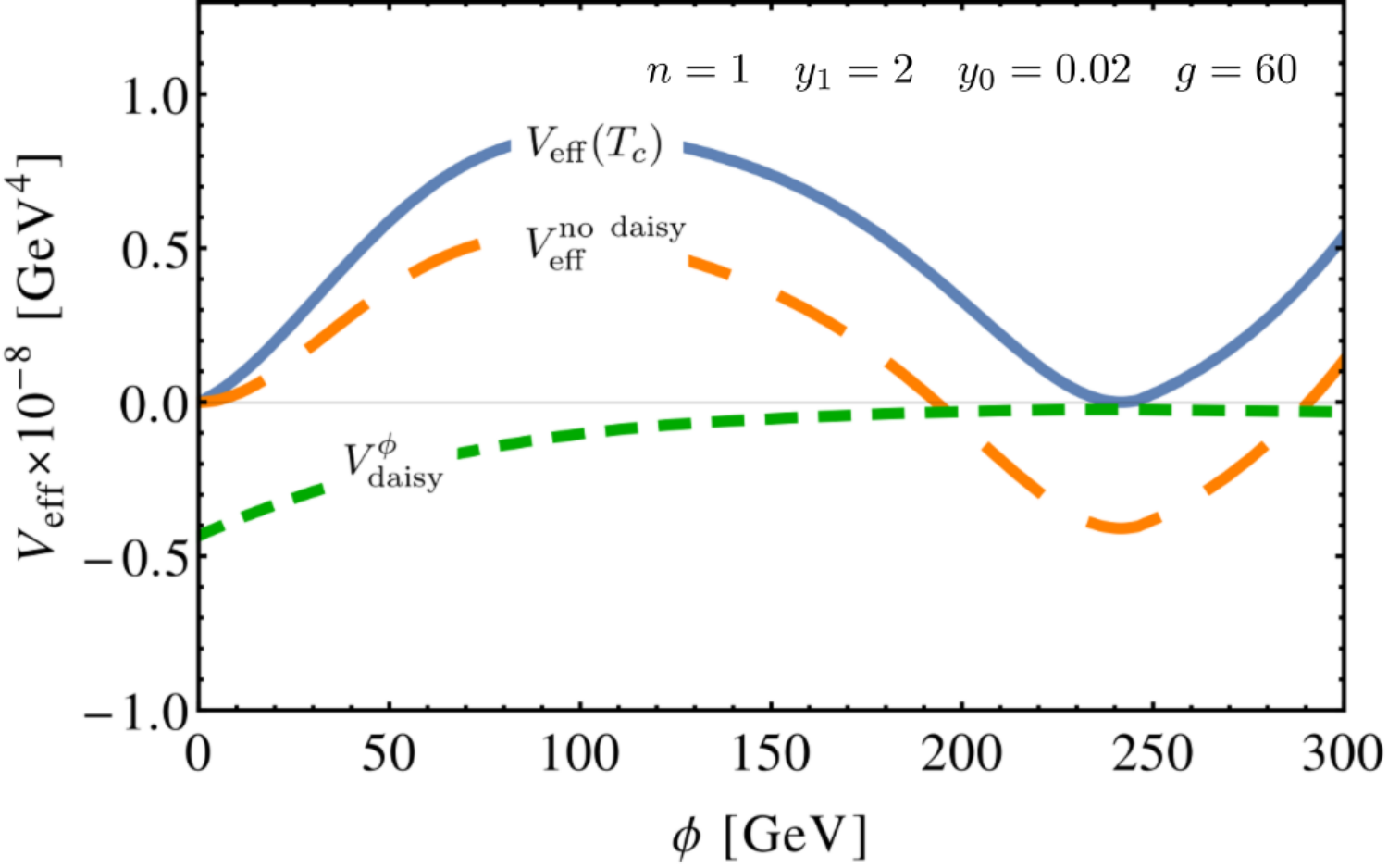}
\end{center}
\caption{Solid line: the full potential at the critical temperature for $y_{1}=2$, $y_{0}=0.02$, $n=1$, $g=60$. Short dashed line: the Higgs boson daisy contribution, showing its effect of lowering the barrier at $\phi=0$ compared to the potential at $\phi\neq0$. This delays and strengthens the phase transition. Long dashed line: the potential without the Higgs boson daisy contribution (but still normalised to 0 at $\phi=0$), showing the phase transition would have occurred earlier without the daisy contribution. Our calculation returns $\phi_{c}=242 \; (240)$ GeV, $T_{c}=111 \; (128)$ GeV with (without) the daisy contribution.}
\label{fig:daisy}
\end{figure}
{\bf 2) Barrier from the $T\neq 0$ one-loop potential}:	
The one-loop finite temperature correction is given by
	\begin{eqnarray}
	V_{1}^{T}(\phi,T)&=& \sum_{i} \frac{g_{i}(-1)^{F}T^{4}}{2\pi^2} \times\\
&&	\int_{0}^{\infty}y^{2}\mathrm{Log}
	\left(1-(-1)^{F}e^{-\sqrt{ y^{2}+{m_{i}^{2}(\phi) }/{ T^{2} } }}\right)\mathrm{d}y.
	\nonumber
	\end{eqnarray}
	We focus on the fermionic contribution, 
	\begin{equation}
        \label{eq:tneq0}
	V^{T}_{f}(\phi,T)=-\frac{gT^{4}}{2\pi^{2}}J_{f}\left(\frac{ m_{f}(\phi)^{2} }{ T^{2} } \right),
	\end{equation}
where $J_{f}(x^{2})$ has a high-temperature expansion for $x^{2} \ll 1$, 
	\begin{equation}
        J_{f}(x^{2}) \approx \frac{7\pi^4}{360}-\frac{\pi^{2}}{24}x^{2}-\frac{x^{4}}{32}\mathrm{Log}\left[\frac{x^{2}}{13.9}\right].
	\end{equation}
The first term in this expansion is constant in $\phi$ and has no effect. The third term is higher order $\sim \phi^{4}$ and can be ignored for the purposes of our discussion here. The second term is crucial as, for decreasing Yukawas, it leads to a barrier between the symmetric and broken phases,
	\begin{equation}
	\label{eq:deltav}
	\delta V \equiv V^{T}_{f}(\phi,T)-V^{T}_{f}(0,T)\approx \frac{gT^{2}\phi^{2}[y(\phi)]^{2}}{96}.
	\end{equation}
This leads to a cubic term in $\phi$, e.g. for $y(\phi) = y_{1}(1-\phi/v)$:	
	\begin{equation}
	\delta V \approx \frac{g y_{1}^{2} \phi^{2} T^{2}}{96} \left(1-2\frac{\phi}{v}+\frac{\phi^{2}}{v^{2}}\right)
	\end{equation}
giving a barrier for the potential. The barrier due to the contribution of Eq.~(\ref{eq:deltav})  is clearly seen in Fig.~\ref{fig:SM_vs_varying}. This is different from the effect noted in \cite{Carena:2004ha} which assumed constant Yukawas. The distinct effect here is that the decreasing Yukawas actually create the barrier on their own as they lead to effectively massless fermions, $m_{f}(\phi) \ll T $, not only around $\phi=0$, but also at $\phi \approx v$. In contrast, for usual mass terms, i.e. linear in $\phi$, only bosonic fields create thermal barriers as the finite-$T$ expansion for bosonic fields contains a cubic term which the fermionic function lacks.\\

{\bf 3) Effects from the Daisy correction}:
The daisy correction comes from resumming the Matsubara zero-modes for the bosonic degrees of freedom:~\cite{Carrington:1991hz,Parwani:1991gq,Arnold:1992rz}
	\begin{equation}
	V_{\rm Daisy}(\phi,T) = \sum_{i}\frac{\overline{g}_{i}T}{12 \pi}\Big \{ m_{i}^{3}(\phi)-\big[m_{i}^{2}(\phi)+\Pi_{i}(T)\big]^{3/2}\Big\}
	\end{equation}
where the sum runs only over scalars and the longitudinal degrees of freedom of the vector bosons ($\bar{g_{i}} \equiv \{ 1,2,1 \}$ for the $\phi$, $W^{\pm}$ and $Z$ bosons) and the $\Pi_{i}$ denote the thermal masses.	

We consider the contribution from the Higgs,
        \begin{equation}
	V^{\phi}_{\rm Daisy}(\phi,T) =\frac{T}{12 \pi}\Big \{ m_{\phi}^{3}(\phi)-\big[m_{\phi}^{2}(\phi)+\Pi_{\phi}(\phi,T)\big]^{3/2}\Big\},
	\end{equation}
where the Higgs boson thermal mass is~\cite{Katz:2014bha}
        \begin{equation}
        \Pi_{\phi}(\phi,T)  =  \left(\frac{3}{16}g_{2}^{2}+\frac{1}{16}g_{Y}^{2}+\frac{\lambda}{2}+\frac{y_{t}^{2}}{4}+\frac{gy(\phi)^2}{48}\right)T^{2}.
        \end{equation}
The novelty is  the dependence of the thermal mass on $\phi$, which comes from 
the $\phi$-dependent Yukawa couplings (these do not enter into the thermal masses for the $W$ and $Z$ bosons at this order). The effect of this term is to lower the effective potential at $\phi=0$, with respect to the broken phase minimum, as long as $\Pi_{\phi}(0,T_{c}) \gg \Pi_{\phi}(\phi_{c},T_{c})$. This is shown in Fig.~\ref{fig:daisy}. By lowering the potential at $\phi=0$, the phase transition is delayed and strengthened. 

\section{Summary}
We have shown how varying Yukawas during the EWPT change the nature of the EWPT due mainly to three effects on the Higgs effective potential:
1) The first effect comes from the $T=0$ one-loop potential. Large Yukawas in the symmetric phase can lead to a significant decrease of the potential in the region $0 < \phi < v$. This can weaken the phase transition.
2) The $T\neq 0$ one-loop contributions from the fermions create a barrier between the $\langle \phi \rangle =0$ and $\langle \phi \rangle  \neq 0$ minima. This can result in a first-order phase transition.
3) Large Yukawas at $\phi \sim 0$ significantly increase the Higgs thermal mass, which, through the Daisy resummation, lowers the potential close to the origin $\phi\sim0$, delaying the phase transition and thereby increasing $\phi_c/T_c$. Note that effect (1) scales as $y_{1}^{4}$, effect (2) as $y_{1}^{2}$ and effect (3) as $y_{1}^{3}$. The net result of these three effects is to give a strong first-order phase transition in large areas of parameter space, while not being disallowed by creating a deeper minimum than the EW one.

The physics of varying Yukawas during the EWPT has important implications for electroweak baryogenesis with rich phenomenology.
{In most explicit realisations, other effects from the new ingredients responsible for the flavour dynamics add up to the ones studied here, which are model-dependent. For instance, in~\cite{Bruggisser:2018mus,Bruggisser:2018mrt}, the nature of the EWPT is mainly controlled by the form of the flavon/dilaton potential, while in Ref.~\cite{Baldes:2016gaf} additional fermions whose mass is also controlled by the flavon/dilaton VEV have a crucial effect. Interestingly, in~\cite{Baldes:2018nel}, SM Yukawa variation as studied here is the dominant effect for making the EW phase transition first-order. 
 Most importantly, in addition to its effects on the nature of the EWPT, this flavour-EW symmetry breaking interplay  has {major} effects on CP violation~\cite{Bruggisser:2017lhc}. }It will be interesting to identify further realistic models and their experimental signatures.


\bibliography{yukewpt} 

\end{document}